\documentclass[11pt,twoside]{pnas-new}
\usepackage{soul,xcolor}
\setstcolor{red}

\newcommand{\rev}[1]{\textcolor{black}{#1}}
\newcommand{\red}[1]{\textcolor{black}{#1}}

\templatetype{pnasresearcharticle} 

\title{Bots increase exposure to negative and inflammatory content in online social systems}

\author[a]{Massimo Stella}
\author[b,*]{Emilio Ferrara} 
\author[a,*]{Manlio De Domenico}

\affil[a]{Fondazione Bruno Kessler, Via Sommarive 18, 38123 Trento, Italy}
\affil[b]{USC Information Sciences Institute, 4676 Admiralty Way 1001, Marina del Rey, CA 90292 USA}

\leadauthor{Stella} 

\authorcontributions{All authors conceived the original study; MDD gathered and processed the data; MS and MDD analysed the data and performed the numerical experiments. All authors wrote the manuscript.}
\authordeclaration{The authors declare no conflict of interest.}
\correspondingauthor{\textsuperscript{*}Corresponding authors. E-mail: mdedomenico@fbk.eu; emiliofe@usc.edu}


\begin{abstract}
Societies are complex systems which tend to polarize into  sub-groups of individuals with dramatically opposite perspectives. This phenomenon is reflected -- and often amplified -- in online social networks where, however, humans are no more the only players, and co-exist alongside with social bots, i.e., software-controlled accounts. Analyzing large-scale social data collected during the Catalan referendum for independence on October~1,~2017, consisting of nearly 4~millions Twitter posts generated by almost 1~million users, we identify the two polarized groups of Independentists and Constitutionalists and quantify the structural and emotional roles played by social bots. We show that bots act from peripheral areas of the social system to target influential humans of both groups, bombarding Independentists with violent contents, \red{increasing their exposure to negative and inflammatory narratives and exacerbating social conflict online. Our findings stress the importance of developing countermeasures to unmask  these forms of automated social manipulation.}
\end{abstract}

\dates{This manuscript was published in the \textit{Proceedings of the National Academy of Sciences 115 (49) 12435-12440} on December 4, 2018}
\doi{\url{www.pnas.org/cgi/doi/10.1073/pnas.1803470115}}

\begin{document}

\verticaladjustment{-2pt}

\maketitle
\thispagestyle{firststyle}
\ifthenelse{\boolean{shortarticle}}{\ifthenelse{\boolean{singlecolumn}}{\abscontentformatted}{\abscontent}}{}

\dropcap{S}ocieties consist of agents engaging in multi-modal social actions with one another in a complex system~\cite{sawyer2005social}. This ``society-as-system'' metaphor inspired many computational studies aimed at identifying, at a microscopic level, how social interactions  might lead to emergent global phenomena such as social segregation~\cite{schelling1971dynamic}, spreading of information~\cite{travers1969experimental} and behavior~\cite{centola2010spread,centola2011experimental}. 
The recent advent of digital communication systems has dramatically shifted the investigation from empirical social interactions in the physical world to online social platforms and technology-mediated interactions~\cite{shirado2013quality}. 
Online platforms revolutionized the ``society-as-system'' metaphor~\cite{lazer2009life} by providing detailed datasets suitable for large-scale investigation of patterns reflecting real-world social phenomena such as the presence and role of influencers in information diffusion~\cite{aral2009distinguishing,onnela2010spontaneous,aral2012identifying,bond201261}, the effect of emotions on social ties~\cite{kramer2014experimental}, or the polarization of agents according to stances~\cite{conover2011political,lee2014social,cruz2014building}. Social media yield an invaluable source of information for learning the mechanisms behind social influence and social dynamics~\cite{gonzalez2011dynamics,borge2016dynamics,bessi2016social}. However, digital systems are not populated only by humans but also by software-controlled agents, better known as bots, programmed to pursue specific tasks, from sending automated messages to assuming specific social or antisocial behaviors~\cite{ferrara2016rise,varol2016online}. Similarly to human interactions, bots might be able to affect structure and function of a social system~\cite{bessi2016social}. 
Understanding how human-bot dynamics drive social behavior is of utmost importance: As postulated by the theory of embodied cognition~\cite{anderson2003embodied}, the presence of robots in a social system affects the way human perceive social norms and how they interact one another and with the robots.

Here, we show how social bots play a central role in the collective dynamics taking place on online social systems during a voting event, namely the Catalan Referendum of October~$1$, 2017. To this end, we monitored the discussion on a popular microblogging platform (Twitter) from September~$22$, 2017 to October~$3$, 2017. We  discovered that bots generated specific content with negative connotation that targeted the most influential individuals among the group of Independentists (i.e., Catalan independence supporters). For our analysis, we first detect bots by using a cutting-edge scalable approach and find that nearly one in three users in this conversation is a bot. 

\section*{Results}

By disentangling the observed social interactions in Retweets (who re-shares the content posted by whom), Replies (who responds to whom) and Mentions (who attracts the attention of whom), we find that humans and bots share similar temporal behavioral patterns in the volume of messages. Both groups display daily excursions resembling a circadian rhythm, with a dramatic increase in  the activity rate on October~$1$. 

\begin{figure*}[ht!]
\centering
\includegraphics[width=12cm]{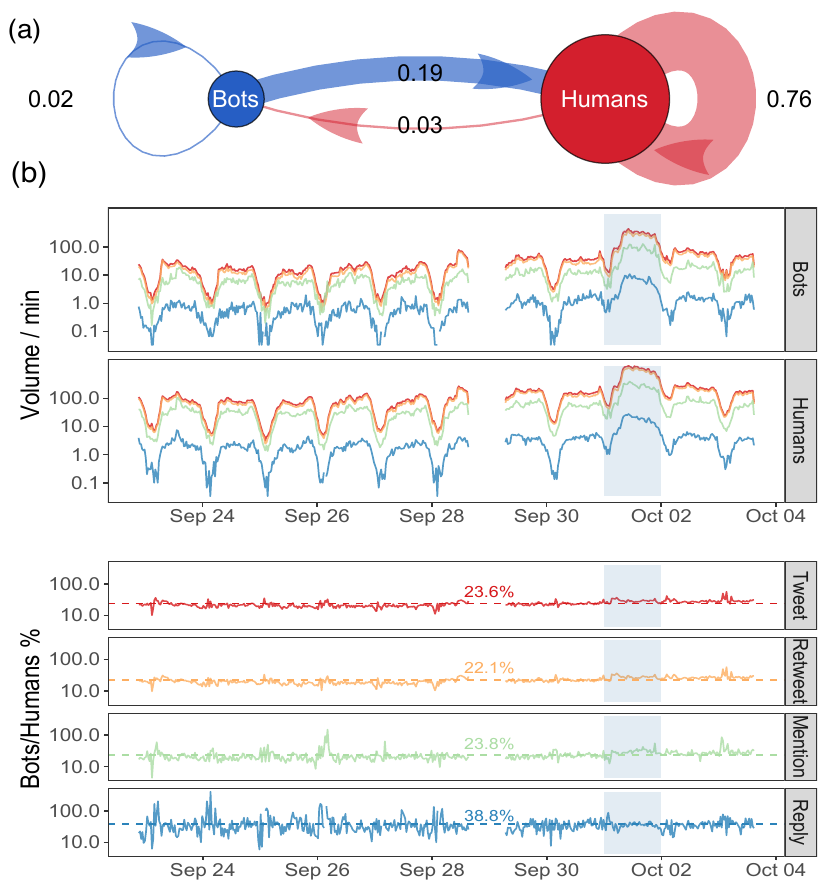}
\caption{\textbf{Social activity of humans and bots over time.} (a): Flowchart of human-bot Twitter interactions across the whole time window. 19\% of the considered interactions are from bots to humans. (b): Upper panel shows the volume per minute for different social actions (Tweet, Retweet, Mention and Reply). Lower panel shows the fraction of volume generated by bots. Shaded areas highlight October 1, 2017, the day of the Catalan referendum.}
\label{fig:socact}
\end{figure*}

Figure~\ref{fig:socact}b (lower panel) shows that bots produced 23.6\% of the total number of posts during the event (Retweets and Mentions show comparable values). Notably, the percentage of Replies generated by bots increases to 38.8\%, suggesting that during this event bots preferred this form of targeted responses.

To better characterize the nature of the observed interactions, we investigate the targets of such intensive social activities. Fig. 1a and Fig.~S1A (SI Appendix) summarize the structure of human-bot interactions. While humans interact mostly with other humans, 19\% of overall interactions are directed from bots to humans mainly through Retweets ($74\%$) and Mentions ($25\%$), Fig.~S1(B--D). 

\begin{figure*}[ht]
\centering
\includegraphics[width=0.85\textwidth]{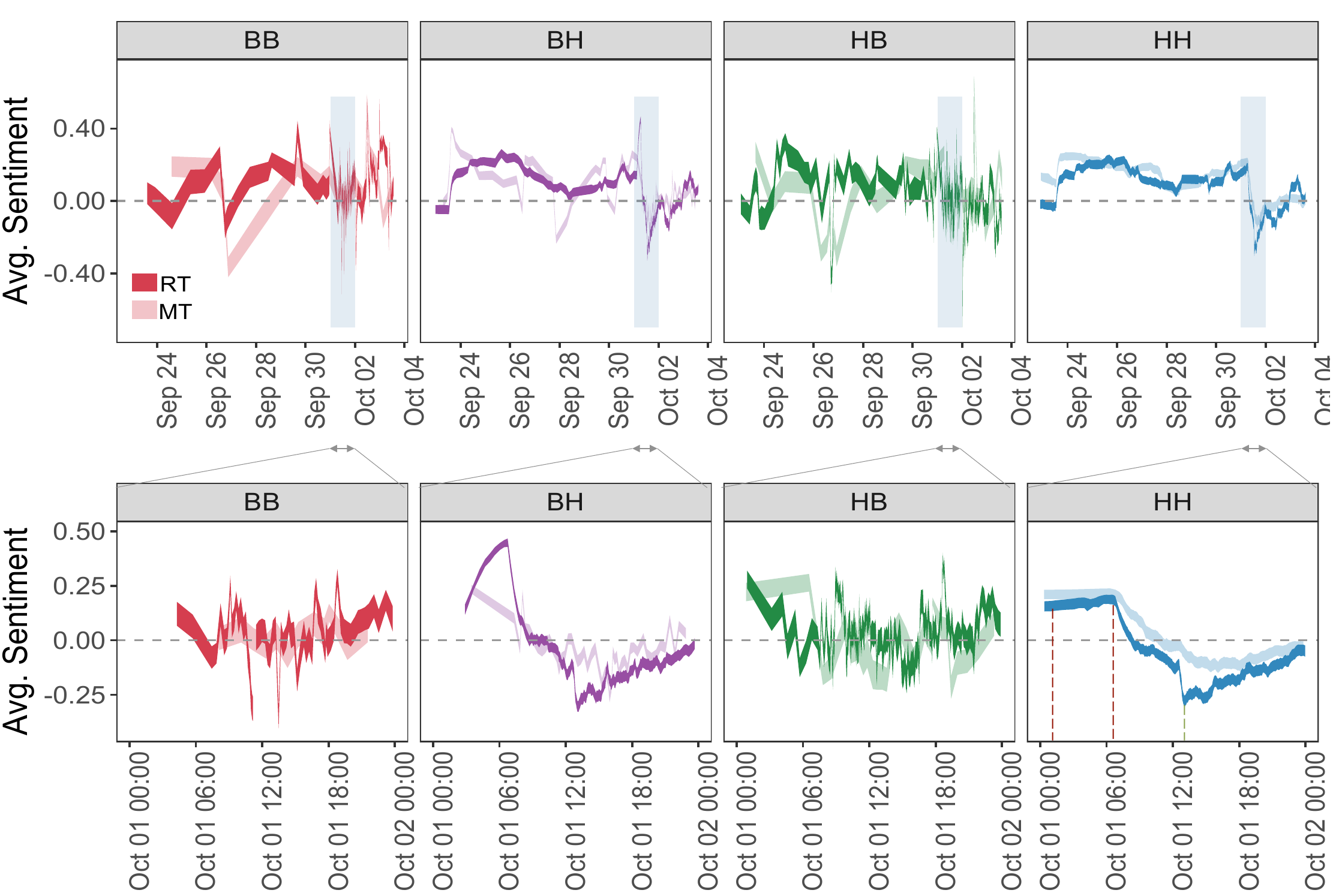}
\caption{\textbf{Sentiment evolution before, during and after the Catalan referendum.} Average sentiment scores for Retweets (full color) and Mentions (lighter colors) over time for human-to-human (HH), human-to-bot (HB), bot-to-human (BH), bot-to-bot (BB). The grey box in the above plots highlights the day before the Catalonia ballot. While bot-to-bot and human-to-bot display no clear trend over time, human interactions display a positive pattern of sentiments until September 30, after which a drop in sentiment up to negative values appears in human-to-human and bot-to-human interactions. In the lower-right sub-panel, negative tweets are generated around 1:00 AM of October 1 but they start spreading only in the morning, after 7:00 AM. Positive tweets start spreading after noon.}
\label{fig:senti}
\end{figure*}

To shed light on the nature of these human-bot interactions we focus on the semantic content of posted messages. Sentiment analysis (see Materials and Methods) reveals interesting differences in emotional trends between humans and bots (Fig.~\ref{fig:senti}). Retweets directed to bots do not display any evident deviation from neutrality (0 sentiment score), while interactions directed towards humans display marked positive and negative trends of sentiment intensity. An analogous behavior happens also for mentions (light colors). These differences indicate that bot-targeted interactions are not significantly influenced by the underlying social dynamics and hence the analysis should focus more on human-targeted interactions (i.e., human-to-human and bot-to-human). 
The sentiment of human-to-human interactions displays marked trends in different phases: (i) a trending positive average sentiment score in the days before September~30 (Fig.~\ref{fig:senti}, HH upper panel); (ii) a sudden drop in sentiment starting from the midnight of October~1 (Fig.~\ref{fig:senti}, HH lower panel) after negative contents start getting reshared; (iii) a peak of negative sentiment on the midday of October~1 and (iv) a later increase in sentiment towards neutrality. These sentiment scores and their related content both indicate that human-to-human interactions are a powerful proxy of the dynamics of underlying real-world social system. The drastic drop of sentiment score from positive to negative among more than 300K human users signals the presence of polarization in the social system, due either to opposing factions exchanging positive/negative messages or to the influence of non-humans. \rev{Figure 2 also highlights important differences between human and bot interactions: The drop in average sentiment evident in bot-to-human interactions is not present in bot-to-bot interactions. This difference indicates that automated content generated and endorsed by bots is not influenced by the social dynamics relative to the referendum: On average, bot-to-bot interactions are not influenced by the human polarization relative to the referendum. Such human polarization is captured by bot-to-human interactions, instead: This distinctiveness indicates that bot-to-human interactions promote human generated content, which is subject to polarization.}

Identifying user polarization, i.e., users being in favour or against a given event or topic, cannot be performed with sentiment only~\cite{taule2017overview}. We overcome this limitation by exploiting a synergy between the network structure of social actions and their emotional intensities, with the aim of identifying stances focused on the voting event in our dataset: Constitutionalists and Independentists to the Catalan referendum. Notice that our network-enhanced stance detection analysis has two major elements of novelty compared to previous approaches~\cite{taule2017overview}, as it not only considers semantic features of messages but also the structure of their exchanges and the nature of their recipients. 

\begin{figure*}[t!]
\centering
\includegraphics[width=16cm]{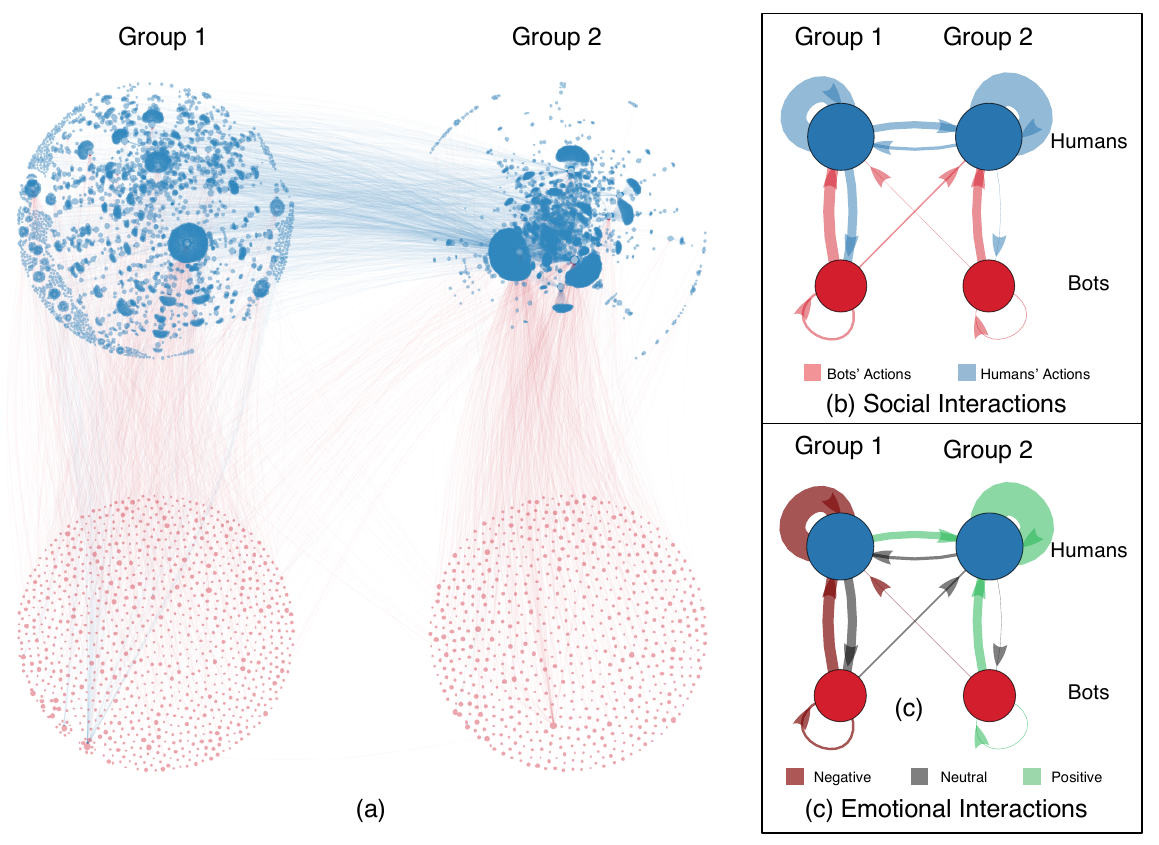}
\caption{\textbf{Network of Twitter interactions.} (a): Visualization of the network among users classified with respect to faction and bot/human class. Nodes indicate users and links encode their social interactions (Retweet and Reply or Mention). Top: sub-networks corresponding to the factions consisting of humans. Bottom: sub-networks of bot factions. Colors encode interactions started by humans (blue) or bots (red). (b): Total traffic of Twitter interactions among humans and bots. Thicker edges indicate higher traffic volume. (b): Median sentiments of Twitter interactions among factions. Interactions with average negative (positive) sentiment are in dark red (green). Black corresponds to interactions on average compatible with neutrality. Distributions of sentiments are tested against neutrality (i.e., 0 sentiment score) with a sign test at a 95\% confidence level. }
\label{fig:interdependency}
\end{figure*}

To capture pivotal trends in the structure of social interactions we focus on the core of the network of social interactions (see Materials and Methods). It is well documented that people tend to retweet each other as a form of social endorsement~\cite{metaxas2015retweets}. To filter out spurious or infrequent interactions, we consider the available multi-modal information and focus on strong social interactions, i.e., those actions where users perform at least a retweet and either a reply or a mention, during the considered time window. We use strong ties to identify the network core, shown in Fig.~\ref{fig:interdependency}. To determine the two underlying polarized groups, we look for a partition that minimizes inter-group interactions and use the Fiedler vector approach~\cite{ding2001min} for an efficient estimation (cf., Methods). The results are shown in Fig.~\ref{fig:interdependency}(A). Each group includes about 6,300 users, with 18\% (12\%) of them being bots in Group~1 (Group~2). Within both groups, human-to-human interactions are the most frequent ones, followed by bot-to-human, see Fig.~\ref{fig:interdependency}(B). Humans in Group~1 direct towards bots almost 100 times more social interactions than in Group~2, suggesting a larger influence of bots on the social dynamics in Group~1 rather than in Group~2. Bot-bot interactions across the two groups are absent since bots mostly interact with humans.

To understand the importance of humans and bots in this network, we calculate the PageRank, a widely used measure of user's importance in online networks~\cite{brin2012reprint}. On average, we find that humans are 1.8 times more central than bots, highlighting that the latter tend to act from the periphery of the social system. Interestingly, despite their peripheral position, bots target their interactions strategically, mostly directing their activity towards human hubs, playing an influential role in the system. If we define the in-degree of a user as the number of its incoming interactions, then the in-degree of humans with respect to interactions incoming only from bots correlates positively with the in-degree with respect to interactions incoming only from humans (Kendall Tau $\kappa \approx 0.62$, p-value~$< 10^{-4}$), indicating that bots tend to target their interactions mainly with the most connected humans. Analogously, also humans tend to interact mainly with the most connected bots (Kendall Tau $\kappa \approx 0.75$, p-value~$< 10^{-4}$). To verify if these effects are genuine, we have performed the same analysis on randomized realizations of the network while preserving the empirical degree distribution. In this test, the observed correlations are no longer present, supporting the hypothesis of a strategic targeting of social interactions. Since hubs in online social networks like Twitter characterize broadcasters and influencers~\cite{aral2012identifying}, the above results suggest that bots interacting with human hubs can influence the social dynamics of both groups, while remaining in the periphery of the microblogging social system. \rev{The volume of bot/human endorsements in the two groups, and the fact that bots mainly target human hubs, indicate that social bots can be influential: They promote human-generated content from hubs, rather than automated tweets, and target significant fractions of human users---as evidenced by the fraction of endorsements shown in Figure 3 (B) and reported in the SI Appendix.}

To harness the emotional structure of the links in the network core, we perform a sentiment analysis of the interactions among humans and bots in the two groups, see Fig.~\ref{fig:interdependency}(C) and Materials and Methods. The resulting atlas of emotional interactions indicates that the average sentiments of human-to-human and bot-to-human interactions are negative within Group 1 and positive within Group 2. This substantial difference in sentiment suggests that the two identified groups endorse their exchanged messages in a different way. In fact, Group 1 preferentially endorses negative content. \rev{The volumes and sentiment polarities reported in Figure~\ref{fig:interdependency}(C) highlight an important mechanism of social contagion played by bots. Firstly, bots direct significant fractions of endorsements to human users, thus actively exposing humans to some type of automatically-generated content. However, this content crucially depends on the targets of the interaction: The polarity of endorsements from bots to humans coincide in both groups with the average sentiment of human-human interactions. In turn, this indicates that bots exploit and promote human-generated content, with the same polarity of the endorsements in a given group of human users. In this way, social bots \red{accentuate the exposure of opposing parties to negative content, with the potential to exacerbate social conflict online.}} 


\begin{figure*}[ht!]
\centering
\includegraphics[width=\textwidth]{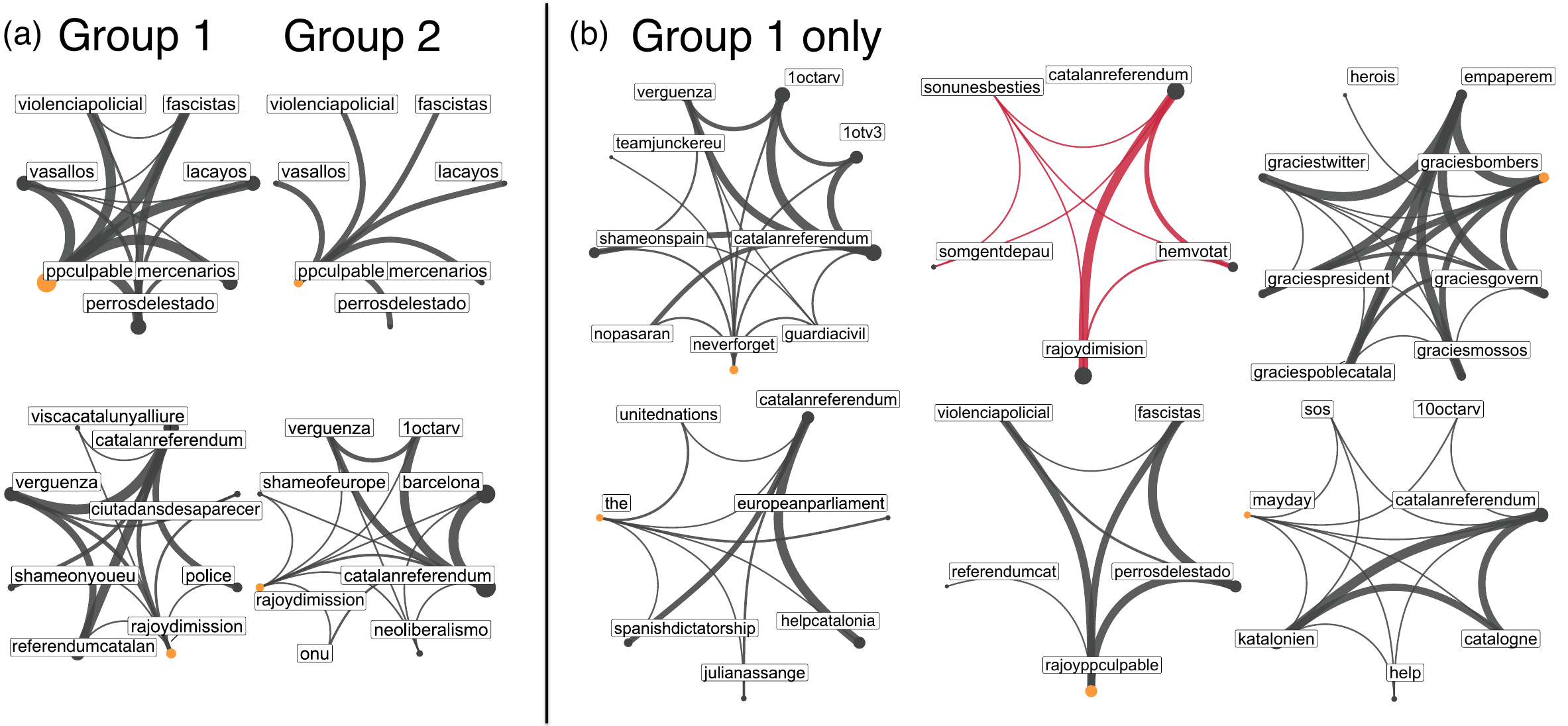}
\caption{\textbf{Hashtag ecosystem reveals group identity.} Hashtags are coupled together if they appear simultaneously in a message, building a network of concepts. Analyzing the hashtag networks obtained from each group, we identify the hashtags which are ranked (a) similarly and (b) very differently in the two groups, to visualize the corresponding neighboring concepts. In (a), low-ranked hashtags coexist in both groups and do not allow to identify the underlying ideology of each group. In (b) top-ranked hashtag that exist only in Group 1 are strongly related to concepts of freedom, independence, fight, shame against the Spanish government, dictatorship and blame against police violence, providing evidence that Group 1 consists of Catalan Independentists. \rev{Remarkably, concepts related to ``sonunesbesties'' (tr. ``they are beasts'') -- highlighted in (b) -- are posted by bots only, whereas the other hashtag networks have contributions by both humans and bots. Note that, for clarity, in panel B we show only hashtags fully characterizing accounts associated with Independentists. Remarkably, concepts related to “sonunesbesties” (tr. “they are beasts”) – highlighted in (b) – are posted by bots only, whereas the other hashtag networks have contributions by both humans and bots.} }
\label{fig:hasheco}
\end{figure*}

To characterize the semantic nature of \rev{group-specific endorsements} (e.g. aggressive, pessimistic, etc.) we build and analyze networks of hashtag co-occurrences (see Materials and Methods), providing a proxy of users' mindset, i.e., the way users perceive and associate concepts \cite{baronchelli2013networks,kenett2014investigating,kenett2018flexibility}. 
A consistency analysis indicates that the two groups post messages about a common set of 4,132 hashtags but associate the corresponding concepts in different ways. Figure~\ref{fig:hasheco} shows how the same hashtags co-occur differently in Group 1 and Group 2. Capitalizing on this finding, we focus on those specific concepts that are most important for one group but most peripheral in the other one. We quantify the importance of concepts by identifying the hashtags with highest degree, strength and closeness centrality---characterizing number of different associations, total frequency of co-occurrences, and how closely hashtags are associated, respectively~\cite{steyvers2005large,baronchelli2013networks,stella2017children}. 

In Group 1, concepts of ``freedom'' and ``independence'' are dramatically associated with ``fight'', ``shame'' against the Spanish government, ``dictatorship'', and blame against ``police violence''. \rev{In Group 2, these associations are completely missing, providing strong quantitative evidence that Group 1 consists of Independentists. By combining this finding with the analysis shown in Fig.~\ref{fig:interdependency}, which highlights the existence of only two groups, we deduce that Group 2 consists of Constitutionalists and non-Independentists.} We further distinguish between associations coming from bots and humans. Negative associations for the content of Group 1 comes exclusively from bots, highlighting their intent of bolstering conflict. 

\rev{To enrich the results provided by our data-driven sentiment and network analyses, we performed human coding of 2,413 tweets posted by humans and social bots (see SI Appendix). The analysis confirms the trends of sentiment polarities for human-to-human interactions, with shared content becoming increasingly pessimistic as a reaction to the violence registered on the onset of the referendum day. Moreover, human coding of the content of automated tweets confirms that bots mainly promoted news-media titles from hubs, mimicking the trend of human emotions and hence boosting sentiments of alarmism, fear and reprobation during and after the vote.}

\section*{Discussion}

Through the synergy of cutting-edge techniques in bot detection, multi-language sentiment analysis, network partitioning and semantic network analysis we find strong evidence of two opposing factions during a large-scale voting event. We provide quantitative findings that the captured online trends in the dataset mirror meaningful events in the real world concerning the voting timeline. Harnessing the structure and the semantic content of social actions within a large-scale dataset, we identify factions as groups of people having opposite stances during the Catalan referendum of October 1, 2017, i.e., Independentists and Constitutionalists. 
Our results demonstrate that bots sustain each faction from the periphery of the online social network structure by mainly targeting human influencers. Bots tend to target human Independentists with messages evoking negative sentiments and associating hashtags with negative connotations. Importantly, we show that bots provide semantic associations, in messages directed to the Independentists, that inspire fight, violence, shame against the government and the police. \rev{In addition to promoting target-specific content generated by human hubs, social bots achieved social contagion also by fabricating automated content within specific communities of humans. The negative associations highlighted in Figure 4 were found only in endorsements relative to Group 1, and were completely absent in messages within Group 2. The specificity of such hatred-inspiring semantic associations provides evidence that bots achieved different social contagion across the Groups also by forging artificial content.}

While software-controlled agents might be beneficial to online networked systems, e.g. by improving the collective performance of human groups~\cite{shirado2017locally}, their improper use can have dramatic effects. Our findings support the hypothesis that bots may influence information diffusion in social media systems~\cite{ferrara2016rise,bessi2016social}, \red{specifically, by accentuating the exposure to negative, hatred-inspiring, inflammatory content, thus exacerbating social conflict online.}
This concerning trend, coupled with the emerging ability to control time-varying networks such as online social systems~\cite{li2017fundamental}, further motivates the crucial need for the development of quantitative techniques like the one proposed here for unmasking the social manipulation enacted by bots.

\paragraph*{Acknowledgements.} 
{\small EF is grateful to the Air Force Office of Scientific Research (award no. FA9550-17-1-0327) for their support.}

\matmethods{\paragraph*{Data collection.}
By following a consolidated strategy, we manually selected a set of hashtags and keywords to collect messages (tweets) posted to a microblogging platform (Twitter). The list contains various general Catalan issue-related terms: $\#Catalunya$, $\#Catalonia$, $\#Catalogna$, $\#1Oct$, $\#votarem$, $\#referendum$, $\#1O$.
We monitored the Twitter stream and collected data by using the Twitter Search API, from September 22, 2017 to past the election day, on October 3, 2017: This allowed us to almost uninterruptedly collect all tweets containing any of the search terms. The data collection infrastructure ran inside FBK servers to ensure resilience and scalability. We chose to use the Twitter Search API (\url{https://dev.twitter.com/rest/public/search}) to make sure that we obtained all tweets that contain the search terms of interest posted during the data collection period, rather than a sample of unfiltered tweets: This precaution avoids incurring in known sampling issues related to collecting data using the Twitter Stream API (\url{https://dev.twitter.com/streaming/overview}) rather than the Search API. 
This procedure yielded a large dataset containing approximately 3.6M unique tweets, posted by 523K unique users. 

\paragraph*{Bot detection.} \rev{Various strategies exist to label social media users as bots or humans~\cite{ferrara2016rise, varol2016online}. Here, we leveraged a scalable and accurate feature-based approach~\cite{ferrara2017disinformation}. Account metadata carry a highly predictive bot signature: We thus identified the top ten most informative account metadata features (see SI Appendix for details). Off-the-shelf learning models were trained on multiple historical ground truth datasets and achieved high detection accuracy (above 90\%) on cross-validation benchmarks. Logistic Regression (LR), our reference model for this study, was selected for its best trade-off between scalability and accuracy: The model is very precise at detecting human accounts---Precision Rate (PR) 98\%, compared to bot accounts (PR: 92\%), while detecting nearly all existing bots---Recall Rate (RR) 99\%, compared to  human users retrieval (RR: 88\%). Furthermore, LR provides binary classifications, rather than continuous probabilistic scores---e.g., like \textit{Botometer} does~\cite{varol2016online}, simplifying the interpretability of resulting annotations without hampering classification accuracy. Finally, random samples of inferred bot and human labels were manually scrutinized for sanity check. All bot detection methods have some inherent limits, e.g., dependency on quality and size of training data, and model generalizability, that we mitigated by employing domain knowledge and state-of-the-art techniques (see SI Appendix for additional discussion).}

\paragraph*{Building the Twitter network.} People from the same faction tend to retweet each other as a form of social endorsement, as documented in the relevant literature \cite{metaxas2015retweets}, while cross-faction retweets are less likely. Considering only retweets would pose the question of how to get rid of spurious or infrequent interactions, possibly by identifying a given retweet threshold. Identifying a threshold would be problematic, as the final network structure might greatly vary with small perturbations on the considered threshold, as it can happen on co-occurrence networks \cite{ninio2014syntactic}. We address this issue by considering strong social interactions: Twitter interactions where users perform at least one retweet but also at least another type of Twitter interaction, be it a mention or a reply during the considered time window. Notice that mentions and replies do not express the same social endorsement of retweets but they can help in identifying the core interactions in the considered social system.
The resulting Twitter Core Network (TCN) included 12 thousands users and 16 thousands directed strong social interactions. Notice that the TCN aggregates together interactions happening over the whole considered time window. However, the frequency of Twitter interactions strongly correlates with the indegree on the TCN (Kendall Tau $\tau = 0.81$), thus indicating that the aggregated network topology is a valid proxy for investigating patterns of Twitter interactions.

\paragraph*{PageRank centrality of humans and bots.}
In the Twitter Core Network we used the average PageRank  \cite{brin2012reprint} as a measure of centrality of human and bot users quantifying how important individual nodes are for information flow in a given network topology. We computed PageRank centrality in Mathematica, which provides normalized values indicating the probability of a random walker to visit a given node. We used 0.85 as dampening factor, as in Google’s Page Rank. On average human users displayed a PageRank of $8.1 \cdot 10^{-4}$ while bots displayed an average PageRank of $4.6 \cdot 10^{-4}$. Hence, on average human users tended to be almost 1.8 times more central than bots in terms of information flow on the Twitter Core Network.

\paragraph*{Network partitioning.}
To detect the two groups in the Twitter Core Network we used the Fiedler vector, a widely used heuristic in spectral graph partitioning \cite{newman2010networks}. The Fiedler vector of a given graph is the eigenvector corresponding to the smallest non-zero eigenvalue (i.e., the algebraic connectivity) of the Laplacian matrix $L=D-A$ of the graph represented by the adjacency matrix $A$ and by the diagonal matrix $D$. Negative and positive entries in the Fiedler vector partition the corresponding network nodes in two sets. One can prove analytically that this heuristic for graph partitioning is a valid approximation for solving the minimum cut problem on general graphs, i.e., partitioning nodes in two groups so that the number of edges across groups is minimized. 
We applied spectral clustering on the undirected version of the TCN and then built randomized partitions. Through direct sampling, we show that the modularity of the Fiedler's partitioning is optimal compared to randomizations even on the original TCN (see SI Appendix).

\paragraph*{Building the hashtag co-occurrence network.}
Hashtags are strings of characters starting with the hash (\#) character and representing the main semantic content of a tweet \cite{tsur2012s}. The literal meaning of hashtags is already considered in the sentiment analysis. Co-occurrence of different hashtags can provide important additional information on the semantic content of tweets, as it was recently shown  \cite{wang2011topic}. Analogously to other association networks in psycholinguistics \cite{kenett2014investigating,kenett2018flexibility}, networks of hashtag co-occurrences represent a powerful proxy of the cognitive profile of users, i.e., the way concepts are perceived and associated by users.
From our Twitter dataset, we build semantic networks of hashtag co-occurrence where nodes represent hashtags and they are linked when co-occurring in at least one tweet. This network definition is in agreement with previous large-scale studies \cite{wang2011topic}. We build one network of hashtag co-occurrences per group. Group 1 (Group 2) co-occurrence network includes 8,451 (7,107) unique hashtags and 29,694 (23,644) links. The two networks overlap for 4,132 hashtags, on which the consistency analysis is performed (see SI Appendix). 

}

\showmatmethods{} 

\bibliography{pnas-sample}

\end{document}